%% file: ms.tex
\documentclass[preprint]{aastex}

\begin{document}

\newcommand{\kms}{\,km\,s$^{-1}$}
\newcommand{\vej}{v_{\mathrm{ej}}}
\newcommand{\lambdao}{\lambda_{\mathrm{o}}}
\newcommand{\lambdar}{\lambda_{\mathrm{r}}}
\newcommand{\zabs}{z_{\mathrm{abs}}}
\newcommand{\zem}{z_{\mathrm{em}}}
\newcommand{\aox}{\alpha_{\mathrm{ox}}}

\newcommand{\rosat}{{\it ROSAT}}

\slugcomment{Draft v2, \today}

\title{Hubble Space Telescope Ultraviolet Spectroscopy of Fourteen Low-Redshift Quasars\footnote{Based
on observations made with the NASA/ESA Hubble Space Telescope, which
is operated by the Association of Universities for Research in
Astronomy, Inc., under NASA contract NAS 5-26555.}}

\shorttitle{HST UV Quasar Spectroscopy}

\shortauthors{Ganguly et al.}

\author{Rajib Ganguly\altaffilmark{2},
        Michael S. Brotherton\altaffilmark{2},
        Nahum Arav\altaffilmark{3},
        Sara R. Heap\altaffilmark{4},
        Lutz Wisotzki\altaffilmark{5},
        Thomas L. Aldcroft\altaffilmark{6},
        Danielle Alloin\altaffilmark{7,8},
        Ehud Behar\altaffilmark{9},
        Gabriela Canalizo\altaffilmark{10},
        D. Michael Crenshaw\altaffilmark{11},
        Martijn de Kool\altaffilmark{12},
        Kenneth Chambers\altaffilmark{13},
        Gerald Cecil\altaffilmark{14},
        Eleni Chatzichristou\altaffilmark{15},
        John Everett\altaffilmark{16,17},
        Jack Gabel\altaffilmark{3},
        C. Martin Gaskell\altaffilmark{18},
        Emmanuel Galliano\altaffilmark{19},
        Richard F. Green\altaffilmark{20},
        Patrick B. Hall\altaffilmark{21},
        Dean C. Hines\altaffilmark{22},
        Vesa T. Junkkarinen\altaffilmark{23},
        Jelle S. Kaastra\altaffilmark{24},
        Mary Elizabeth Kaiser\altaffilmark{25},
        Demosthenes Kazanas\altaffilmark{4},
        Arieh Konigl\altaffilmark{26},
        Kirk T. Korista\altaffilmark{27},
        Gerard A. Kriss\altaffilmark{28},
        Ari Laor\altaffilmark{9},
        Karen M. Leighly\altaffilmark{29},
        Smita Mathur\altaffilmark{30},
        Patrick Ogle\altaffilmark{31},
        Daniel Proga\altaffilmark{32},
        Bassem Sabra\altaffilmark{33},
        Ran Sivron\altaffilmark{34},
        Stephanie Snedden\altaffilmark{35},
        Randal Telfer\altaffilmark{36},
        Marianne Vestergaard\altaffilmark{37}
}
\altaffiltext{2}{Department of Physics \& Astronomy, University of
Wyoming (Dept. 3905), 1000 East University Avenue, Laramie, WY
82071; ganguly@uwyo.edu}
\altaffiltext{3}{Center for Astrophysics and Space Astronomy,
University of Colorado, 389 UCB, Boulder, CO 80309-0389}
\altaffiltext{4}{Laboratory of Astronomy and Solar Physics, NASA
Goddard Space Flight Center, Greenbelt, MD 20771}
\altaffiltext{5}{Astrophysikalisches Institut Potsdam, An der
Sternwarte 16, 14482 Potsdam, Germany}
\altaffiltext{6}{Harvard-Smithsonian Center for Astrophysics, 60
Garden Street, Cambridge, MA 02138}
\altaffiltext{7}{ESO, Alonso de Cordova 3107, Vitacura, Casilla
19001, Santiago 19, Chile}
\altaffiltext{8}{UMR 7158, CEA-CNRS-Universit\'e Paris 7,
DSM/DAPNIA/Service d'Astrophysique, CEA/Saclay, France}
\altaffiltext{9}{Department of Physics, Technion, Haifa 32000,
Israel}
\altaffiltext{10}{Department of Physics, University of California,
Riverside, CA 92521}
\altaffiltext{11}{Department of Physics and Astronomy, Georgia State
University, Astronomy Offices, One Park Place South SE, Suite 700,
Atlanta, GA 30303}
\altaffiltext{12}{Research School of Astronomy and Astrophysics
(RSAA), Mount Stromlo Observatory, Cotter Road, Weston ACT 2611,
Australia}
\altaffiltext{13}{Institute for Astronomy, 2680 Woodlawn Drive,
Honolulu, HI 96822-1897, USA}
\altaffiltext{14}{Department of Physics and Astronomy, University of
North Carolina, Chapel Hill, NC 27599}
\altaffiltext{15}{Institute of Astronomy and Astrophysics - National
Observatory of Athens, I. Metaxa \& Vas. Pavlou, Palea Penteli,
25236 Athens, Greece}
\altaffiltext{16}{Canadian Institute of Theoretical Astrophysics,
University of Toronto, 60 Saint George Street, Toronto, ON M5S 3H8,
Canada}
\altaffiltext{17}{Departments of Astronomy and Physics, and Center
for Magentic Self-Organization, University of Wisconsin, Madison, WI
53703}
\altaffiltext{18}{Department of Physics and Astronomy, University of
Nebraska, Lincoln, NE 68588-0111}
\altaffiltext{19}{Department of Physics, University of California,
Davis, CA 95616.}
\altaffiltext{20}{Large Binocular Telescope Observatory, 933 N.
Cherry Street, Tucson, AZ, 85721-0065}
\altaffiltext{21}{Department of Physics \& Astronomy, York
University, 4700 Keele St., Toronto, Ontario, M3J 1P3, Canada}
\altaffiltext{22}{Space Science Institute, 4750 Walnut Street, Suite
205, Boulder, CO 80301}
\altaffiltext{23}{Center for Astrophysics and Space Sciences 0424,
University of California, San Diego, CA 92093}
\altaffiltext{24}{SRON Netherlands Institute for Space Research,
Sorbonnelaan 2, 3584 CA Utrecht, The Netherlands}
\altaffiltext{25}{Department of Physics and Astronomy, Johns Hopkins
University, 3400 North Charles Street, Baltimore, MD 21218}
\altaffiltext{26}{Department of Astronomy and Astrophysics, and
Enrico Fermi Institute, University of Chicago, 5640 South Ellis
Avenue, Chicago, IL 60637}
\altaffiltext{27}{Department of Physics, Western Michigan
University, Kalamazoo, MI 49008-5252}
\altaffiltext{28}{Space Telescope Science Institute, 3700 San Martin
Drive, Baltimore, MD 21218}
\altaffiltext{29}{Homer L. Dodge Department of Physics and
Astronomy, University of Oklahoma, 440 West Brooks Street, Norman,
OK 73019}
\altaffiltext{30}{Department of Astronomy, Ohio State University,
140 West 18th Avenue, Columbus, OH 43210}
\altaffiltext{31}{Spitzer Science Center, California Institute of
Technology, 1200 East California Boulevard, Pasadena, CA 91125}
\altaffiltext{32}{Department of Physics, University of Nevada, Las
Vegas, NV 89154}
\altaffiltext{33}{Faculty of Natural and Applied Sciences, Notre
Dame University, Zouk Mosbeh, Lebanon}
\altaffiltext{34}{Department of Physics, Baker University, P.O. Box
65, Baldwin City, Kansas 66006-0065 }
\altaffiltext{35}{Apache Point Observatory, 2001 Apache Point Road,
P.O. Box 59, Sunspot, NM 88349-0059}
\altaffiltext{36}{Orbital Sciences}
\altaffiltext{37}{Steward Observatory, The University of Arizona,
933 Cherry Ave., Tucson, AZ 85721}
%
%%%%%%%%%%%%%%%%%%%%%%%%%%%%%%%%%%%%%%%%%%%%%%%%%%%%%%%%%%%%%%%%%%%%
\begin{abstract}
We present low-resolution ultraviolet spectra of 14 low redshift
($\zem \lesssim 0.8$) quasars observed with HST/STIS as part of a
Snap project to understand the relationship between quasar outflows
and luminosity. By design, all observations cover the \ion{C}{4}\
emission line. Nine of the quasars are from the Hamburg-ESO catalog,
three are from the Palomar-Green catalog, and one is from the Parkes
catalog. The sample contains a few interesting quasars including two
broad absorption line (BAL) quasars (HE\,0143-3535, HE\,0436-2614),
one quasar with a mini-BAL (HE\,1105-0746), and one quasar with
associated narrow absorption (HE\,0409-5004). These BAL quasars are
among the brightest known (though not the most luminous) since they
lie at $\zem<0.8$. We compare the properties of these BAL quasars to
the $\zem<0.5$\ Palomar-Green and $\zem>1.4$\ Large Bright Quasar
samples. By design, our objects sample luminosities in between these
two surveys, and our four absorbed objects are consistent with the
$v \sim L^{0.62}$\ relation derived by \citet{lb02}. Another quasar,
HE\,0441-2826, contains extremely weak emission lines and our
spectrum is consistent with a simple power-law continuum. The quasar
is radio-loud, but has a steep spectral index and a lobe-dominated
morphology, which argues against it being a blazar. The unusual
spectrum of this quasar resembles the spectra of the quasars
PG\,1407+265, SDSS\,J1136+0242, and PKS\,1004+13 for which several
possible explanations have been entertained.
\end{abstract}


%%%%%%%%%%%%%%%%%%%%%%%%%%%%%%%%%%%%%%%%%%%%%%%%%%%%%%%%%%%%%%%%%%%%
\section{Introduction}

Outflows from active galactic nuclei (AGN) come in many
observational classes. Seyfert galaxies show blue shifted UV and
X-ray absorption lines hundreds of {\kms} wide \citep{cren99}, while
the UV troughs of quasar outflows can span tens of thousands of
{\kms} as manifested in broad absorption line (BAL) quasars
\citep{lynds67,weymann85,turn88}. One of the current driving
questions in the AGN field is what is the connection, if any,
between the intrinsic luminosity of an AGN and the kinematic
properties of the outflow (e.g., terminal velocity and
velocity-width of the observed trough).

Radiative acceleration, thought to be the principal driver of such
outflows, predicts that the terminal velocity should scale as $v
\sim L^n$ where $0.25 < n < 0.5$ \citep*{alb94}. Qualitatively, such
a progression is likely to exist given the observed fact that
outflow in Seyfert galaxies terminate at $\sim 1000$\,\kms, while
the BAL outflows extend out to $\sim 30,000$\,\kms. However, the
quantitative trend is unclear given the lack of objects in between
these populations. From an analysis of $\sim 56$\ archived {\it
Hubble Space Telescope} (HST) and {\it International Ultraviolet
Explorer} (IUE) spectra of $z< 0.5$\ quasars from the Palomar-Green
survey, \citet{lb02}\ showed that such a trend may indeed exist. The
soft X-ray weak (SXW, defined as $\aox<-2.0$) quasars, which exhibit
BALs in their UV spectra, show a relation of the form $v \sim
L^{0.62 \pm 0.08}$.

This is a higher power-law index than predicted and implies that the
radiation-pressure force multiplier has a luminosity dependence
\citep{lb02}. The force multiplier \citep*{cak75} is factor that
expresses the sum total effect of all lines and edges in
transferring momentum from the incident spectrum of photons to the
gas. The luminosity dependence of the force multiplier arises from a
variety of sources like the dust content and ionization state of the
gas \citep[e.g., ][]{alb94,mur95,sulentic06}. Indeed, considerations
of accretion disk-winds by \citet{mur95} in both the context of the
broad emission-lines and broad absorption-lines require the presence
of shielding gas that prevents X-rays from over-ionizing the
outflowing gas. More recent considerations of the force multiplier
show that it is also sensitive to black-hole mass \citep{pk04},
which then favors a steeper index than predicted by \citet{alb94}.
The \citet{lb02} index is also consistent with the prediction from
\citet*{psd98} for AGN with $L/L_{Edd} \geq M_{\mathrm{max}}^{-1}$,
where $M_{\mathrm{max}}$\ is the maximum value of the force
multiplier.

Another crucial issue is whether the fraction of objects that show
intrinsic absorption is luminosity dependent. Here we are facing a
large gap between the established statistics for BAL quasars (see
below) and the results from a biased sample of Seyferts which showed
that 10 out of the available 17 HST/UV spectra of Seyferts show
intrinsic absorption \citep{cren99}. \citet{kriss02} find a that a
similar fraction of Seyferts (16/34) exhibit absorption in the
\ion{O}{6} $\lambda\lambda$1031.926,1037.617 doublet. At
low-redshift, \citet{gan01a} found about $\sim25$\% of quasars from
the HST {\it Quasar Absorption Line Key Project} appear to show
absorption at $\zabs \sim \zem$ (i.e., ``associated'' absorption),
comparable to the study of \citet{lb02} for the $z < 0.5$\
Palomar-Green quasars. However, it must be noted that not all
absorption found near quasars is necessarily intrinsic, and
moreover, not necessarily part of an outflow. A follow-up study of
low-redshift quasars with associated absorption showed that only
$\sim30$\% were time-variable \citep{wise04}. Furthermore, large
statistical studies of absorption in quasars, like the HST {\it
Quasar Absorption Line Key Project}, tend to be biased toward UV
bright targets in order to obtain adequate signal-to-noise spectra.

For high-redshift quasars ($z > 1.4$) intrinsic absorption (usually
BALs) has been studied in large samples using ground-based
telescopes. The Large Bright Quasar Survey \citep{lbqs6} members at
these redshifts have $M_\mathrm{V} = -26$\ to $-28$\ and an
intrinsic fraction of BAL quasars, in the redshift range $1.5 < z <
3.0$, of $22\pm4$\% \citep{hf03}. The FIRST Bright Quasar Survey
\citep{second,third} members have $M_\mathrm{V} = -26$\ to $-29$\
and a BAL incidence of 18\% \citep{becker00}. \citet{ves03} find
that about 25\% of quasars show evidence for intrinsic, low-velocity
\ion{C}{4} absorption from a careful consideration of possible
contamination from intervening systems in a heterogenous, yet large,
sample of moderate redshift ($1.5 < z < 3.0$) quasars. In addition,
\citet{rich01b} find that, in \ion{C}{4} absorption systems observed
toward $z\sim2.5$\ FIRST quasars at large velocity separations, as
many as 30\% may be intrinsic to the background quasars. Such
systems must clearly arise in a high-velocity outflow.

The relationship between the velocity of AGN outflows and their
intrinsic luminosity, as well as the fraction of objects that show
outflows, are poorly known due to two simple selection effects. The
vast majority of AGN outflows are identified by detecting absorption
features associated with the \ion{C}{4} $\lambda\lambda$1548.20,
1550.77 doublet. From ground-based observations this line is only
observed in objects with redshift $z > 1.3$\ and therefore the
population observed tends to be the higher luminosity quasars, of
which $\sim$10\% are BAL quasars. In order to observe objects with
$z < 1.3$\ we must use space-based UV observatories. Only $\sim$20
of the UV brightest AGN which were observed with the HST (out of
$\sim$70) show evidence for outflows, compared with upwards of 500
BAL quasars identified in ground-based observations. The statistics
are especially poor for the luminosity range $10^{44}$\,erg
s$^{-1}$\ to $10^{45.5}$\,erg s$^{-1}$ (corresponding roughly to the
absolute magnitude range $-21.9 \lesssim M_\mathrm{V} \lesssim
-25.6$).

To fill in this dearth of data, we proposed an HST Snap project to
obtain low-dispersion STIS UV spectra of intermediate-luminosity
AGN. Unfortunately, the low efficiency of this program only yielded
observations of fourteen objects out of 200 approved targets, and we
present the data here. Several of the objects are of individual
interest based on their absorption properties. In the next section
(\S\ref{sec:data}), we present the spectra obtained and discuss our
data reduction. In \S\ref{sec:results}, we present our results and
compare them with those from other quasar samples. We briefly
summarize our findings in \S\ref{sec:summary}.

%%%%%%%%%%%%%%%%%%%%%%%%%%%%%%%%%%%%%%%%%%%%%%%%%%%%%%%%%%%%%%%%%%%%
\section{Data}
\label{sec:data}

Our HST/STIS \citep{stis98,kimble98} observations were carried out
using the G230L grating and the 52\arcsec$\times$0.5\arcsec slit
which provides spectra over the wavelength range 1570-3180\,\AA\ at
a dispersion of 1.58\,\AA/pixel (and a 2 pixel per resolution
element sampling rate). For these Snap observations, we used
exposure times of 900s with three exceptions.. We used an exposure
time of 720s seconds for the quasars HE\,0354-5500, HE\,0436-2614,
and PG\,1435-067. (These three quasars were part of the bright end
of our initial list of 200 targets, defined as objects with $B<16$,
and therefore did not require a full 900s exposure.)

We used the standard pipeline which provides fully reduced and
calibrated spectra. According to the STIS Instrument Handbook
\citep{stis}, data reduced using the pipeline have the following
calibration uncertainties: 0.5-1.0 pixels (0.79--1.58\,\AA) in
absolute wavelength calibration, and 4\% (0.02 dex) in absolute
spectrophotometry. The S/N for our spectra were typically $\sim$14
per pixel at 2000\,\AA\ and 3000\,\AA.

\begin{figure}
% Scaling for two-column preprint mode
\epsscale{1.0}
% Scaling for manuscript mode
\epsscale{0.5}
\plotone{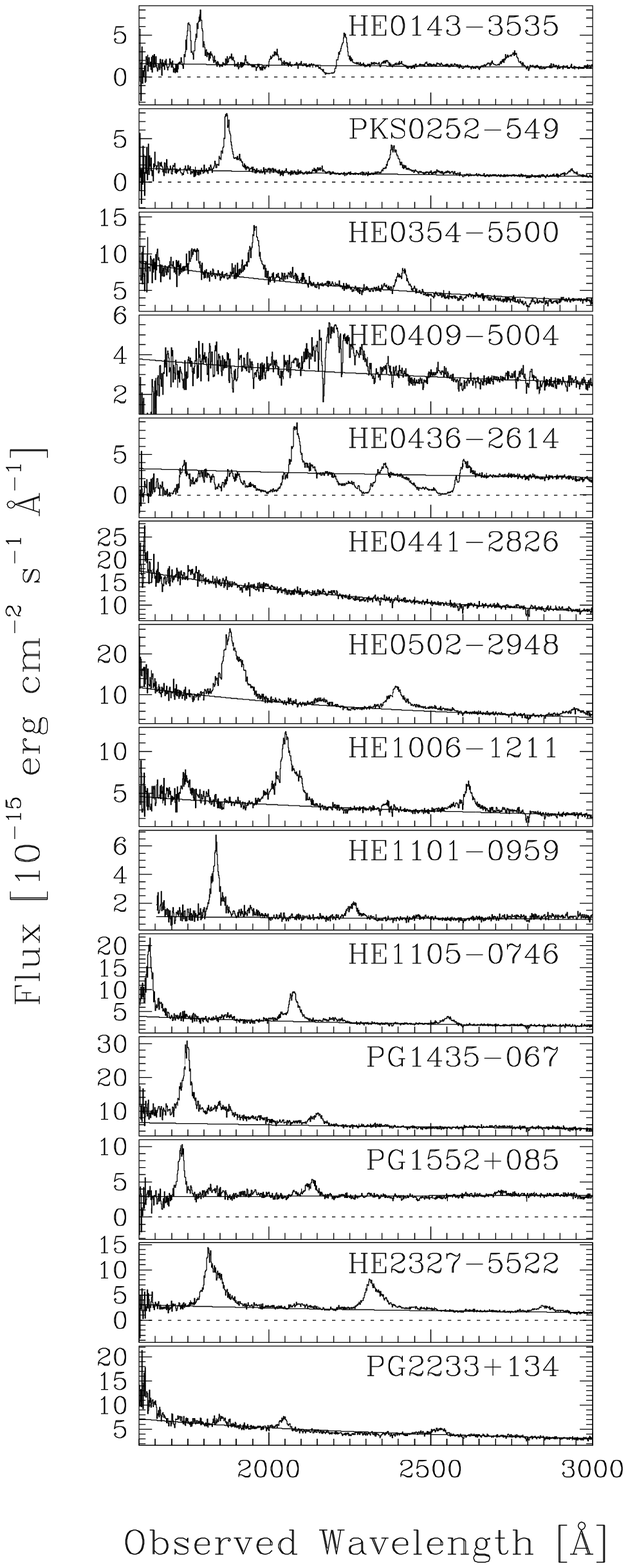}
\protect\caption{We show the spectra (flux versus observed
wavelength, solid histogram) for the subsample of quasars observed
for our Snap program. The spectra are order by right ascension. In
addition, we overlay the power-law fit (smooth curve) described in
\S\ref{sec:results}.}
\label{fig:stisdata}
\end{figure}

Figure~\ref{fig:stisdata} shows the spectra obtained for our
program. The data are publicly available both at the Multi-mission
Archive at Space Telescope\\ (MAST:
http://archive.stsci.edu/index.html, {\sc fits} files only) and at
the University of Wyoming AGN Research Group web site
(http://physics.uwyo.edu/agn, both {\sc fits} and {\sc ascii}
files).

%%%%%%%%%%%%%%%%%%%%%%%%%%%%%%%%%%%%%%%%%%%%%%%%%%%%%%%%%%%%%%%%%%%%
\section{Results}
\label{sec:results}

\subsection{Sample Characteristics}

To characterize this sample and place it in context with larger
samples of low-redshift quasars, we carried out power-law fits to
all spectra with the goal of computing UV spectral slopes and
luminosities. The fits were carried out using an arbitrary number of
superposed Gaussian to mimic the contribution from emission lines.
That is, we fit the following functional form to our spectra:
\begin{equation}
F_\lambda = F_{\lambdao} \left [ \left ( \lambda \over \lambdao
\right )^\alpha + \sum_{i=1}^{m} w_i \exp \left ( - {{(\lambda -
\lambda_i)} \over {\sigma_i}} \right )^2 \right ], \label{eq:fit}
\end{equation}
where $F_{\lambdao}$\ is the normalizing flux at reference wavelength
$\lambdao$, $\alpha$\ is the spectral index, $m$\ is the number of
emission-line components each with a relative strength $w_i$, width
$\sigma_i$, and centered at wavelength $\lambda_i$. The best-fit was
determined using the Numerical Recipes Marquardt-Levenburg
$\chi^2$-minimization routines \citep{nrpress}, and optimal number of
Gaussian components was determined using an F-test. In carrying out
the fits, we omitted regions that were clearly, or potentially,
affected by absorption (as subjectively determined by RG). The results
of the power-law fits are listed in Table~\ref{tab:qsoprops} and
overlayed on the observed spectra in Figure~\ref{fig:stisdata}. For
uniformity, we used a common (observer's frame) reference wavelength,
$\lambdao=1800$\,\AA\ for all fits. Table~\ref{tab:qsoprops} lists the
quasar name (column 1), quasar redshift ($z$, column 2), continuum
flux density at 1800\,\AA\ ($F_{\lambdao}$, column 3), continuum
power-law index ($\alpha$, column 4, with the sign convention
$F_\lambda \sim \lambda^\alpha$), the luminosity at rest-frame
wavelength $\lambdar=3000$\,\AA\ ($L_{\lambdar}$, column 5), and the
\ion{C}{4} emission-line equivalent width for unabsorbed quasars
(column 6). We note that in a few cases (e.g, PG 2233+134) where the
spectra cover wavelengths redward of the \ion{Al}{3} + \ion{C}{3}]
emission line, the continuum fit can be artificially elevated due to
the presence of \ion{Fe}{2-III} emission \citep{vw01}. The
observer-frame flux density at observed wavelength $\lambdao$\ was
converted to rest-frame luminosity at rest-wavelength $\lambdar$\ via:
\begin{eqnarray}
\lambdar L_{\lambdar} & = & 4 \pi D_L(z)^2 (1+z) \lambdar F_{\lambdar} \\
                      & = & 4 \pi D_L(z)^2 (1+z) \lambdar F'_{\lambdar (1+z)}
                                           \nonumber \\
                      & = & 4 \pi D_L(z)^2 (1+z) \lambdar F'_{\lambdao}
                      \left [ {\lambdar(1+z)} \over {\lambdao}
                      \right ]^\alpha, \nonumber
\label{eq:lum}
\end{eqnarray}
where $D_L(z)$\ is the luminosity distance to the quasar, and we have
used the continuum portion of the eq.~\ref{eq:fit} in the second
substitution. For clarity, we have used $F'$\ (primed) to indicate the
flux in the observer-frame (which is the flux derived using
eq.~\ref{eq:fit}), and $F$\ (unprimed) to indicate the flux in the
rest-frame. The luminosity distance was computed using the listed
quasar redshift and a $\Omega_M=0.27$, $\Omega_\Lambda=0.73$,
H$_\mathrm{o}=71$\,\kms Mpc$^{-1}$\ cosmology. The mean luminosity of
the sample is $\langle \lambda L_\lambda(3000\,\mathrm{\AA}) \rangle =
(7.76 \pm 0.02) \times 10^{44}$\,erg s$^{-1}$, and has a range
spanning (0.3--31.4) $\times 10^{45}$\,erg s$^{-1}$. Our fits imply a
mean spectral index of $\langle \alpha_{\mathrm{UV}} \rangle = -1.13
\pm 0.01$, with a standard deviation in the distribution of
$\sigma_\alpha = 0.53$.

\begin{figure}
% Scaling for two-column preprint mode
\epsscale{1.0}
% Scaling for manuscript mode
\epsscale{0.5}
\plotone{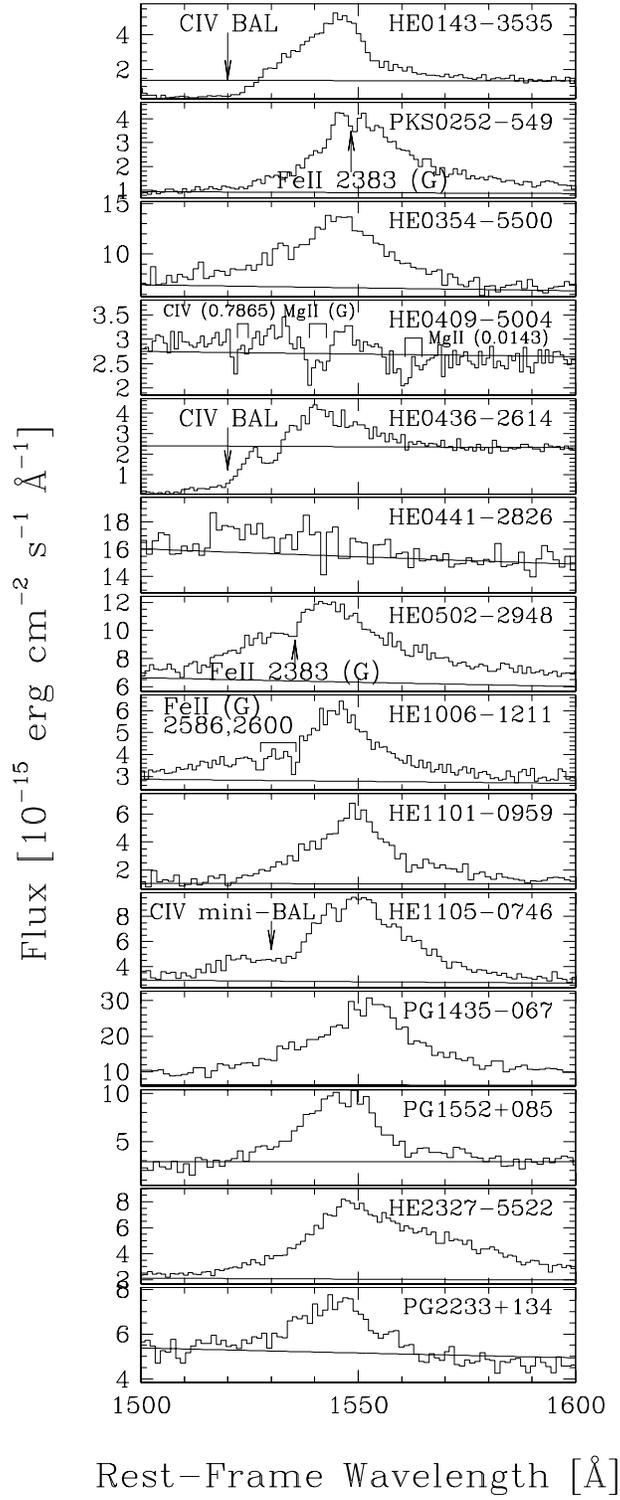}
\protect\caption{For the fourteen quasars in this sample, we show
the region around the \ion{C}{4} emission line. From top to bottom,
the quasars are ordered in increasing redshift. Absorption features
that appear in this wavelength range are marked, and the best-fit
power-law is also shown as a smooth curve.}
\label{fig:civem}
\end{figure}

In Figure~\ref{fig:civem}, we show the region around the \ion{C}{4}
emission line for all 14 quasars. Several quasars show absorption on
top of the emission line, but most of these are identified with
Galactic lines from \ion{Fe}{2} and \ion{Mg}{2} as indicated in the
figure. Four quasars do appear to show intrinsic or associated
\ion{C}{4} absorption and we discuss these in the following
sections.

%%%%%%%%%%%%%%%%%%%%%%%%%%%%%%%%%%%%%%%%%%%%%%%%%%%%%%%%%%%%%%%%%%%%
\subsection{Absorbed Quasars}
\label{sec:absqsos}

Four of the 14 quasar spectra appear to show intrinsic/associated
\ion{C}{4} absorption: HE 0143-3535, HE 0409-5004, HE 0436-2614, and
HE 1105-0746. Figure~\ref{fig:bals} shows the spectra as luminosity
versus rest-wavelength, and ordered (top to bottom) by decreasing
maximum ejection velocity of absorption. To characterize the
absorption properties of these quasars, we carried out fits to the
\ion{C}{4} emission line. We took the same approach as in the
previous section, and the fits are also shown (with the $1\sigma$\
statistical uncertainty) in the figure. These fits allowed us to
compute the following properties which are listed in
Table~\ref{tab:absprops}: Balnicity Index \citep[column
2;][]{weymann91}, Intrinsic Absorption Index \citep[column
3;][]{hallai}, \ion{C}{4} absorption equivalent width (column 4),
Maximum velocity of absorption (column 5).

We note that the fits use the minimum number of Gaussian components
and, as a result, there may be some systematic uncertainties. In
particular, for the two BAL quasars, HE 0143-3535 and HE 0436-2614,
if there is significant asymmetry in the intrinsic emission-line
profiles \citep[e.g., ][and references
therein]{wills93,bro94,rich02b}, it is possible that our fits do not
fully recover the correct shape. Parameters such as balinicity index
that depend on the power-law fit only should not be affected by the
systematics of trying to reproduce the emission-line profile.

%%%%%%%%%%%%%%%%%%%%%%%%%%%%%%%%%%%%%%%%%%%%%%%%%%%%%%%%%%%%%%%%%%%%
\subsection{Comparison To Other Samples}

\begin{figure}
% Scaling for two-column preprint mode
\epsscale{1.0}
% Scaling for manuscript mode
\epsscale{0.75}
%
% Color figure for electronic edition only
%\rotatebox{-90}{\plotone{f4_color.eps}}
% BW figure for paper edition
\rotatebox{-90}{\plotone{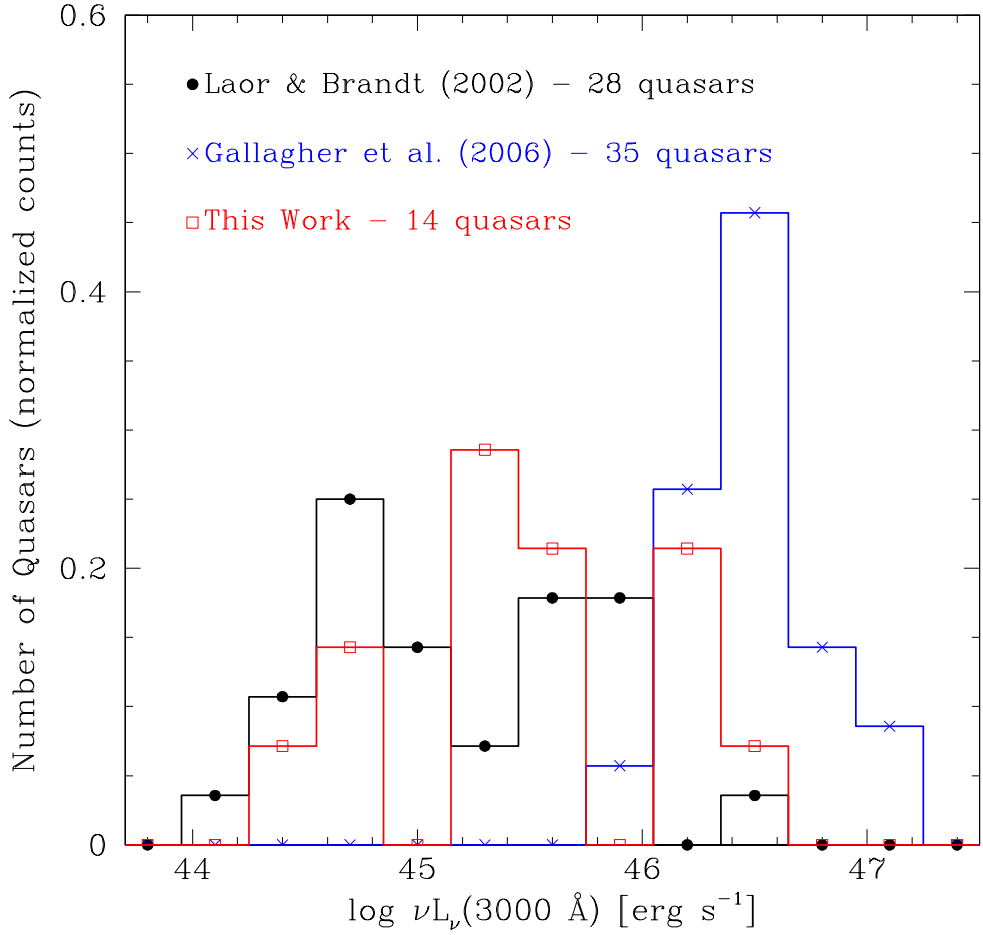}}
\protect\caption{We compare
the luminosity distribution of quasars in our sample to that of the
BQS and LBQS samples. Data for the BQS was taken from \citet{lb02}.
Data from the LBQS was taken from \citet{gallsc06}.}
\label{fig:lumdist}
\end{figure}

Figure~\ref{fig:lumdist} shows a comparison of the quasar
luminosities from this sample to $\zem<0.5$ quasars in the Bright
Quasar Survey \citep*[BQS,][]{brandt,bg92,sg83} and $\zem > 1.4$\
broad absorption-line quasars from the Large Bright Quasar Survey
\citep[LBQS,][]{gallsc06,hf03,lbqs6}. For the LBQS, we restrict the
comparison of the luminosity distribution to the subsample of
absorbed quasars from \citet{gallsc06} since this is the most
interesting aspect of our initial survey. While the entire LBQS
sample does extend down to $\zem=0.2$, good ground-based
spectroscopic observations of \ion{C}{4} spectral region are
available only for $\zem > 1.4$ where the UV doublet shifts into the
optical band. The luminosities of our quasar sample lie in between
these two surveys, though with some overlap with the low-redshift
BQS.

\begin{figure}
% Scaling for two-column preprint mode
\epsscale{1.0}
% Scaling for manuscript mode
\epsscale{0.75}
%
% Color figure for electronic edition only
%\rotatebox{-90}{\plotone{f5_color.eps}}
% BW figure for paper edition
\rotatebox{-90}{\plotone{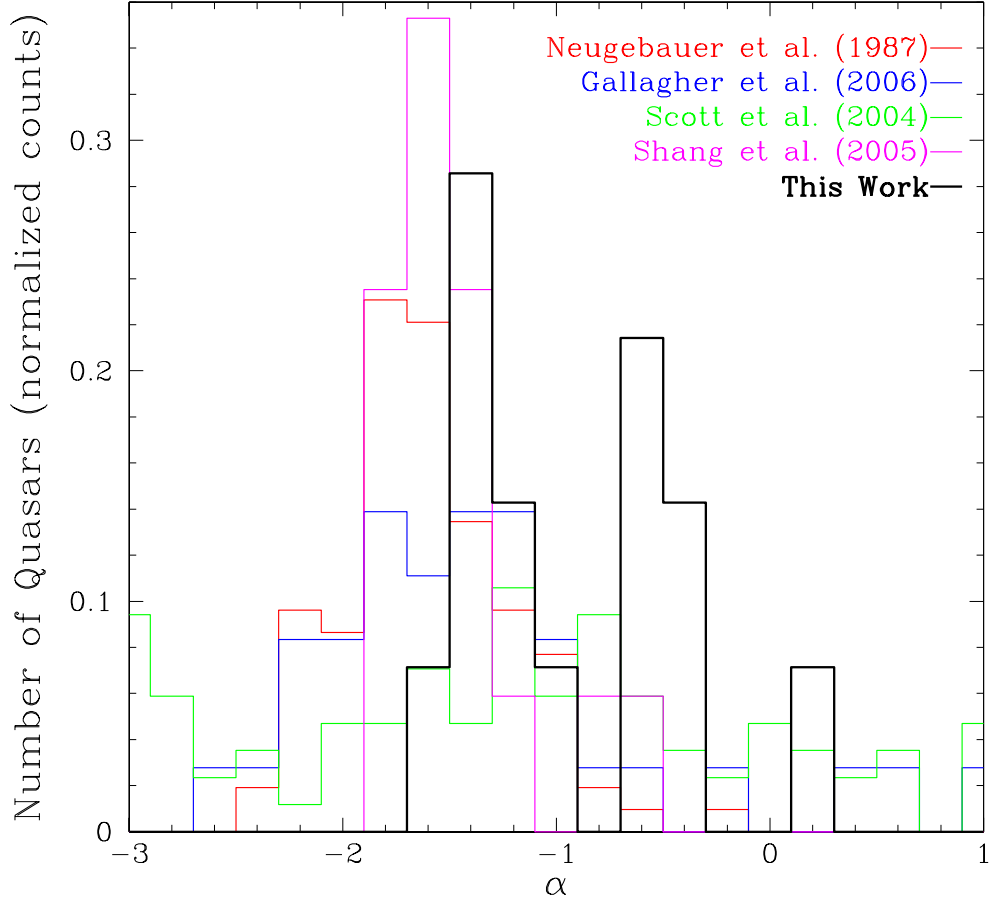}}
\protect\caption{In the above panel, we show the distribution of UV
spectral indices ($F_\lambda \sim \lambda^\alpha$) for our sample.
The histogram is normalized to unit area. We compare this
distribution to other samples including: Palomar-Green quasars from
\citet{neu87}, Large Bright Quasar Survey BAL quasars from
\citet{gallsc06}, quasars used in the {\it FUSE} composite by
\citet{scott04b}, and quasars from \cite{shang05}.}
\label{fig:plaw}
\end{figure}

Figure~\ref{fig:plaw} shows a comparison of our quasar spectral
indices to several other samples: Palomar-Green quasars from
\citet{neu87}, Large Bright Quasar Survey BAL quasars from
\citet{gallsc06}, quasars used in the {\it FUSE} composite by
\citet{scott04b}, and quasars from \citet{shang05}. In general,
there is agreement between the comparison samples (and in particular
between the Palomar-Green, absorbed LBQS, and \citet{shang05}
samples), with a peak near $\alpha \sim -1.8$. \citet{scott04b} show
a correlation between the spectral index and luminosity, with more
luminous objects having bluer spectra. This correlation nicely
explains the difference in shape between the FUSE composite and the
EUV portion of the HST composite spectrum from \citet{telfer02}. The
spectral indices derived for the quasars in our samples tend to be
redder than those samples, with a peak near $\alpha \sim -1.4$, and
an aforementioned mean of $-1.14$.

One potential explanation for this is the effect of reddening.
Galactic reddening is an unlikely explanation, as the largest color
excess in our sample is E($B-V$)$=0.063$\ toward PG $1435-067$, and HE
$1006-1211$. \citet{bl05} note that quasars (particularly from the
BQS) appear redder (as measured by a two-point spectral index between
1549\,\AA\ and 4861\,\AA) as more intrinsic \ion{C}{4} absorption is
present, implying that intrinsic dust is present along sight-lines
that also produce intrinsic \ion{C}{4} absorption. While intrinsic
reddening may be important \citep{gaskell04,gaskell06}, it is unlikely
to be the source of reddening in our quasars, as our three reddest
quasars (PG\,1552+085, HE\,1101-0959, and PG\,1435-067) do not show
intrinsic \ion{C}{4} absorption. In addition, our quasars are
generally redder than the LBQS BAL quasars from \citet{gallsc06},
although this may arise from the luminosity effect described by
\citet{scott04b}.

\begin{figure}
% Scaling for two-column preprint mode
\epsscale{1.0}
% Scaling for manuscript mode
\epsscale{0.75}
%
% Color figure for electronic edition only
%\rotatebox{-90}{\plotone{f6_color.eps}}
% BW figure for paper edition
\rotatebox{-90}{\plotone{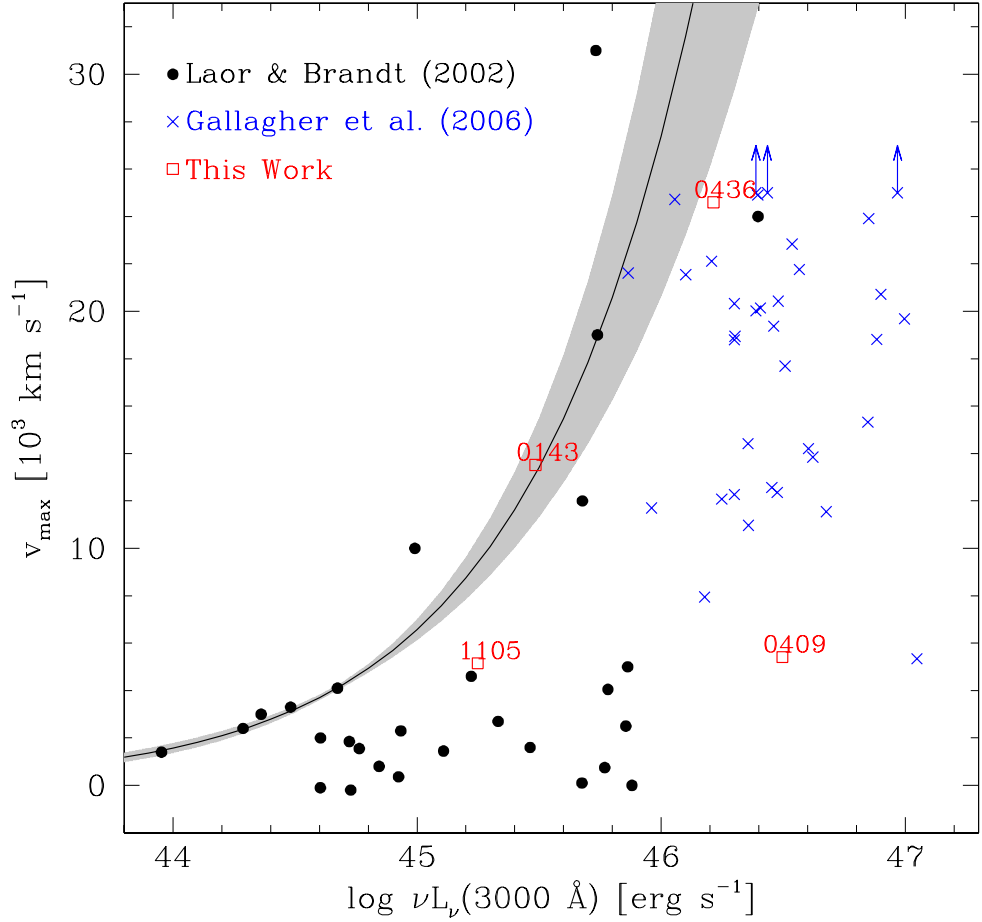}}
\protect\caption{We reproduce the maximum velocity of absorption
versus 3000\,\AA\ luminosity plot from \citet[][revised for a
concordance cosmology]{lb02}. The solid curve is their least-squares
fit for the soft X-ray weak quasars in the $z<0.5$ Bright Quasar
Survey (see eq.~\ref{eq:lb}, which has also been corrected for
cosmology). The shaded region about the curve is the 1$\sigma$\
confidence uncertainty. We overplot the four absorbed quasars from
this work.}
\label{fig:lbrev}
\end{figure}

Since one of our initial goals was to test the radiative-driving
hypothesis for quasar winds, we show in Figure~\ref{fig:lbrev} a
plot of the maximum velocity of absorption against luminosity for
our absorbed quasars, the BQS quasars from \cite{lb02}, and the LBQS
BAL quasars from \citet{gallsc06}. From a consideration of soft
X-ray weak quasars in the BQS sample, \citet{lb02} reported an
apparent envelope to the maximum velocity of absorption as a
function of luminosity \citep[specifically $M_\mathrm{V}$, but see
also][for a plot versus 2500\,\AA\ luminosity]{gallsc06}. We
reproduce the \citet{lb02} plot, revised for a concordance
cosmology, with the four quasars from this work also shown. The two
BAL quasars from this sample, HE\,0143-3535 and HE\,0436-2614, both
seem to lie close to the (cosmology corrected) best-fit curve
derived by \citet{lb02}:
\begin{equation}
v_\mathrm{max} = 4100\,\mathrm{km~s}^{-1}\, \left [ {{\nu
L_\nu(2500\mathrm{\AA})} \over {10^{44.67} \mathrm{erg~s}^{-1}}}
\right ]^{0.62}. \label{eq:lb}
\end{equation}
The other two absorbed quasars, like most of the absorbed BQS
quasars lie below this curve. There are several potential
explanations for this. Since the measured velocities only measure
the line-of-sight component and not the transverse (i.e., in the
plane of the sky) component, the scatter below this curve is
possibly explained by orientation effects. In addition, other
driving mechanisms, such as magnetocentrifugal accretion disk winds
\citep{everett05} or thermally-driven winds \citep{kk01}, may be
important for absorbers at lower velocities. At very low velocities,
the absorbers may not even arise from an outflowing wind. They may
arise from satellite galaxies around the quasar host galaxies, or
even from the interstellar medium of the host galaxy
\citep[e.g.,][]{gan06a,ham01}.

%%%%%%%%%%%%%%%%%%%%%%%%%%%%%%%%%%%%%%%%%%%%%%%%%%%%%%%%%%%%%%%%%%%%
\subsection{Notes on Individual Quasars}

\begin{figure*}
% Scaling for two-column preprint mode
\epsscale{1.05}
% Scaling for manuscript mode
\epsscale{0.5}
% Color figure for electronic edition only
%\rotatebox{-90}{\plotone{f3_color.eps}}
% BW figure for paper edition
\rotatebox{-90}{\plotone{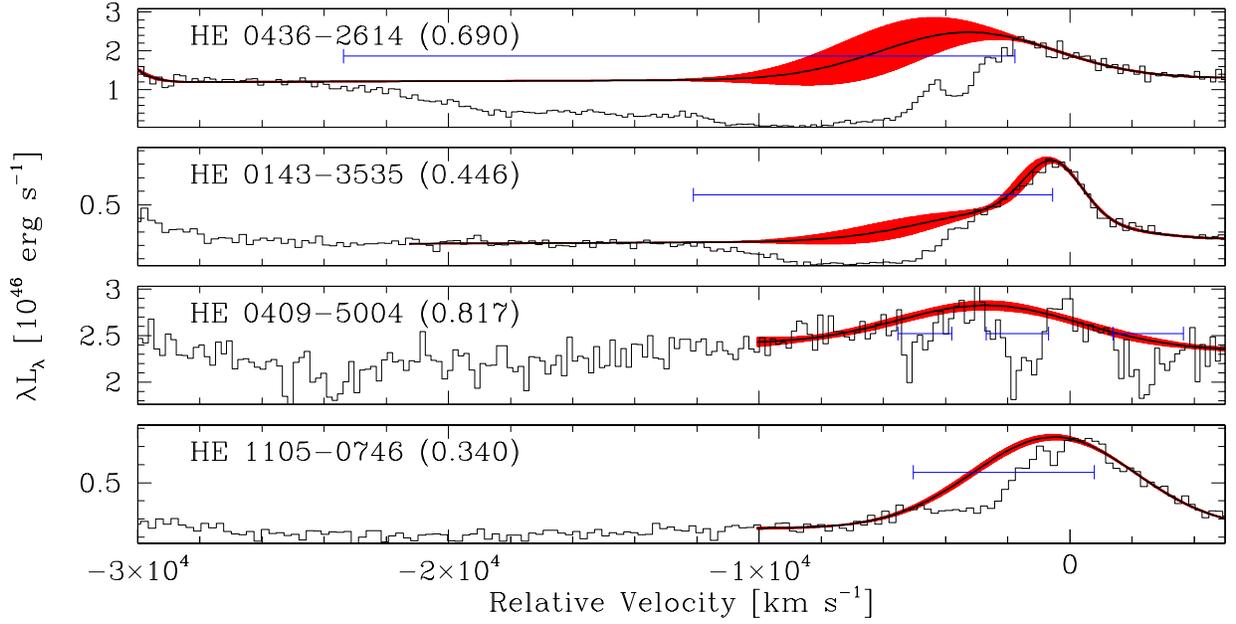}}
\protect\caption{In the above panel, we show the spectra (luminosity
versus relative velocity, with $\lambda=$ 1548.204\,\AA\ $(1+\zem)$
defining the velocity zero-point) of the \ion{C}{4} emission and
absorption lines for the four AGN that show intrinsic/associated
absorption. From top to bottom, the AGN spectra are shown in order
of decreasing maximum ejection velocity of absorption. The
luminosity profiles are shown as a black histogram. The smooth curve
around the \ion{C}{4} emission line is our assessment of the
effective continuum level (i.e., power-law continuum plus emission
lines), with the shaded region indicating the $1\sigma$\
uncertainty. The horizontal bars indicate wavelength regions that
were omitted from the fit due to the presence of absorption.}
\label{fig:bals}
\end{figure*}

{\noindent\bf HE 0143-3535:} This is one of two BAL quasars in the
sample. The BAL is detected clearly in both \ion{C}{4} and
\ion{N}{5}. It is marginally detected in \ion{H}{1} Ly-$\alpha$ in a
relatively low signal-to-noise ($S/N \approx 6$) region of the
spectrum. Reproduction of the \ion{C}{4} emission-line for this
quasar is problematic (Figure~\ref{fig:bals}). Our best-fit requires
two components, a strong narrow component and a broad weaker
component, as motivated by an inflection on the red wing that cannot
be produced by a single Gaussian component. The components are
offset by 1441\,\kms, implying significant asymmetry. The
uncertainty in the shape of the emission-line, and consequently the
uncertainty in the \ion{C}{4} absorption equivalent width is
significant. Regardless, if our assessment of the \ion{C}{4}
absorption equivalent width is accurate, then this quasar may be
significantly absorbed in the soft X-ray band. Furthermore, if the
shape of the \ion{C}{4} emission line is accurate, then the
absorption profile is consistent (within errors) of occulting the UV
continuum only.

%%%%%%%%%%%%%%%%%%%%%%%%%%%%%%%%%%%%%%%%%%%%%%%%%%%%%%%%%%%%%%%%%%%%%

{\noindent\bf HE\,0409-5004:} Three absorption doublets appear on
the weak \ion{C}{4} emission-line of this quasar, which is well-fit
by a single Gaussian. Two of those are the \ion{Mg}{2}
$\lambda\lambda$2796,2803 doublet at $\zabs=-0.00006$\ (i.e.,
Galactic absorption) and $\zabs=0.0143$. The third doublet, which
appears blueward of the other two, is an associated narrow
\ion{C}{4} absorber at $\zabs=0.7865$\ with a velocity offset of
$4579 \pm 165$\,\kms\ from the emission redshift of the quasar. The
absorption-line system is also detected in \ion{H}{1} Ly-$\alpha$.
The \ion{N}{5} $\lambda\lambda$1238, 1242 and \ion{O}{6}
$\lambda\lambda$1031, 1038 doublets are in our wavelength range, but
neither is detected. We do detect absorption from low-ionization
species, \ion{C}{2}, \ion{S}{2-III}, the Lyman-$\beta$ line, and
possibly the Lyman limit. This implies that the gas is in a
relatively low-ionization state (compared to other associated narrow
absorption-line systems). Thus, it is possible that this system,
while close in velocity, is not physically linked with the quasar
central engine.

%%%%%%%%%%%%%%%%%%%%%%%%%%%%%%%%%%%%%%%%%%%%%%%%%%%%%%%%%%%%%%%%%%%%%

{\noindent\bf HE\,0436-2614:} This quasar is the second in the
sample that hosts a BAL. The BAL absorption in this quasar in very
strong. The wavelength range of our spectrum gives us coverage down
to the \ion{O}{6} $\lambda\lambda$1031, 1038 doublet. We detect
broad absorption from \ion{C}{4}, \ion{Si}{4}, \ion{N}{5},
\ion{H}{1} Ly-$\alpha$, \ion{P}{5}, and \ion{O}{6}. The detection of
\ion{P}{5}, which has a low relative abundance, implies a
combination of high column density in the flow and, perhaps, high
metallicity \citep{ham98}. In the \ion{C}{4} BAL profile, there
appears to be a curious transition in the apparent strength of the
absorption (whether from coverage fraction or column density
effects) around $v=-1.2 \times 10^4$\,\kms\ where the emission line
appears to terminate (Figure~\ref{fig:bals}). Redward of this
velocity (i.e., closer to the emission line), the absorption appears
to be completely saturated with full coverage of both the continuum
and broad emission-line regions. In the velocity range $-1.5 \times
10^4 \lesssim v$[\kms]$ \lesssim -1.2 \times 10^4$, the absorption
appears to a have a flat bottom with a normalized flux of $\sim
0.31$\ (in units of the continuum $+$ emission-line flux). In the
velocity range $-1.8 \times 10^4 \lesssim v$[\kms]$ \lesssim -1.5
\times 10^4$, the profile is also flat-bottomed, but with a
normalized flux of $\sim 0.38$. Blueward of $v \sim -1.8 \times
10^4$\,\kms, the absorption tapers off (either due to coverage
fraction change or column density change) to no apparent absorption
at $\sim -2.4 \times 10^4$\,\kms.

%%%%%%%%%%%%%%%%%%%%%%%%%%%%%%%%%%%%%%%%%%%%%%%%%%%%%%%%%%%%%%%%%%%%%

{\noindent\bf HE\,0441-2826:} The spectrum of this quasar is
peculiar. The \ion{C}{4} emission-line is very weak ($\sim10$\% of
the continuum strength, the lowest of the sample), and the spectrum
is consistent with a simple power-law with Galactic \ion{Fe}{2} and
\ion{Mg}{2} absorption. [The redshift of the quasar is known from
optical emission lines, so we are secure in the identification of
this UV line as \ion{C}{4}.] In the rest-frame of the quasar, the
spectrum covers the wavelength range 1376--2726\,\AA. This range
also covers the \ion{Si}{4} $\lambda\lambda$1393, 1402 doublet, the
\ion{Al}{3} $\lambda\lambda$1855, 1863 + \ion{C}{3}] $\lambda$1909
blend and some of the \ion{Fe}{2} UV multiplet, but these are not
significantly detected. The quasar is detected in the NRAO-VLA Sky
Survey \citep{nvss} with a 1.4 GHz flux of $147.8 \pm 4.5$\,mJy, and
a radio core fraction of $\sim 0.24 \pm 0.01$. The quasar is also
listed in the Parkes-MIT-NRAO survey catalog (PMN J$0443-2820$) with
a 4.85 GHz flux of $56 \pm 11$\,mJy. We conclude that the quasar has
a radio-loudness \citep{kel94} of $\log R^* = 1.6 \pm 0.2$, and a
radio spectral index of $\beta = -0.78 \pm 0.16$ ($F_\nu \sim
\nu^\beta$). While a weak/absent emission lines may be indicative of
a blazar, the marginal radio-loudness, steep radio spectral index,
and lobe-dominated morphology seem to rule this out. Alternatively,
this quasar could be in the class of unusual emission-line objects
that includes the radio-quiet PG\,1407+265
\citep[$\zem=0.94$;][]{mcdowell95}, the radio-loud SDSS\,J1136+0242
\citep[$\zem=2.4917$;][]{hall04}, and possibly the intrinsic
spectrum of the radio-loud BAL PKS\,1004+13
\citep[$\zem=0.24$;][]{wbl99}. These objects have very weak/absent
and highly-blueshifted high-ionization emission lines.

%%%%%%%%%%%%%%%%%%%%%%%%%%%%%%%%%%%%%%%%%%%%%%%%%%%%%%%%%%%%%%%%%%%%%

{\noindent\bf HE\,1105-0746:} Our spectrum of this quasar covers the
rest-frame wavelength range 1187--2350\,\AA, and the \ion{H}{1}
Ly-$\alpha$ + \ion{N}{5}, \ion{Si}{4}, \ion{C}{4}, \ion{He}{2}, and
\ion{Al}{3} + \ion{C}{3}] emission lines are clearly visible. The
\ion{C}{4} emission-line is well described by a single Gaussian, and
part of the blue wing of the \ion{C}{4} emission-line of this quasar
appears to be absorbed. This absorption lies above the continuum and
therefore the quasar has a zero balnicity. However, the width of the
absorption is broader than the \ion{C}{4} doublet separation, so
this is an example of a so-called mini-BAL. While the absorption
does not dip below the continuum, the maximum depth of the
absorption is equal to the strength of the continuum. That is, this
mini-BAL is consistent with saturated absorption of the continuum,
and no absorption of the \ion{C}{4} broad emission line. This has
been observed before in BALs \citep[e.g.,][]{arav99b}. Curiously, we
do not detect mini-BAL absorption in \ion{N}{5} or \ion{Si}{4}. The
lack of \ion{Si}{4} absorption may be an ionization effect. The lack
of \ion{N}{5} absorption may be due to a combination of the
noisiness of the spectrum in the wavelength region and dilution by
the strong \ion{H}{1} Ly-$\alpha$\ line if indeed the absorber does
not occult the broad emission-line region.

%%%%%%%%%%%%%%%%%%%%%%%%%%%%%%%%%%%%%%%%%%%%%%%%%%%%%%%%%%%%%%%%%%%%

{\noindent\bf PG\,1552+085:} Based on an {\it IUE} spectrum,
\citet{turn97} claimed that this quasar was host to a BAL. However,
a more recent analysis by \citet{sulentic06} of the HST/STIS-G230L
spectrum presented here yield a balnicity index of $\approx 0$. The
putative location of the \ion{C}{4} BAL through lies in a low
signal-to-noise region of the spectrum and it is difficult to
ascertain if the continuum blueward of the \ion{C}{4} broad
emission-line is absorbed.

%%%%%%%%%%%%%%%%%%%%%%%%%%%%%%%%%%%%%%%%%%%%%%%%%%%%%%%%%%%%%%%%%%%%
\section{Summary}
\label{sec:summary}

While our initial survey was designed to fill in the statistical gap
between luminous quasars and Seyfert galaxies to further understand
the relationship between outflows and luminosity, the low-efficiency
of the program did not make this feasible. Out of the approved list
of 200 targets, only fourteen objects were observed. The
observations of fourteen nearby bright AGN resulted in the discovery
of a few interesting objects which merit further study: two new,
bright BAL quasars, one new mini-BAL quasar, and one quasar with
unusually weak emission lines. Our observations are in agreement
with $v_\mathrm{max}$-luminosity relation reported by \citet{lb02},
and do lie in between the BQS and LBQS samples.

\acknowledgements

Support for this work was provided by NASA through grant number {\em
HST}-GO-09507, from the Space Telescope Science Institute, which is
operated by the Association of Universities for Research in
Astronomy, Inc., under NASA contract NAS 5-26555.

%%%%%%%%%%%%%%%%%%%%%%%%%%%%%%%%%%%%%%%%%%%%%%%%%%%%%%%%%%%%%%%%%%%%
%\clearpage
%\bibliographystyle{apj}
%\bibliography{bibliography}

%%%%%%%%%%%%%%%%%%%%%%%%%%%%%%%%%%%%%%%%%%%%%%%%%%%%%%%%%%%%%%%%%%%%
\clearpage
%%%%%%%%%%%%%%%%%%%%%%%%%%%%%%%%%%%%%%%%%%%%%%%%%%%%%%%%%%%%%%%%%%%%
%%%%%%%%%%%%%%%%%%%%%%%%%%%%%%%%%%%%%%%%%%%%%%%%%%%%%%%%%%%%%%%%%%%%
\clearpage
%%%%%%%%%%%%%%%%%%%%%%%%%%%%%%%%%%%%%%%%%%%%%%%%%%%%%%%%%%%%%%%%%%%%
\clearpage
%%%%%%%%%%%%%%%%%%%%%%%%%%%%%%%%%%%%%%%%%%%%%%%%%%%%%%%%%%%%%%%%%%%%
%\clearpage
%%%%%%%%%%%%%%%%%%%%%%%%%%%%%%%%%%%%%%%%%%%%%%%%%%%%%%%%%%%%%%%%%%%%
\clearpage
%%%%%%%%%%%%%%%%%%%%%%%%%%%%%%%%%%%%%%%%%%%%%%%%%%%%%%%%%%%%%%%%%%%%

\input{tab1.tex}
\input{tab2.tex}

\end{document}

%% file: tab1.tex
\begin{deluxetable}{lccrr@{\,$\pm$\,}lr@{\,$\pm$\,}l}
\tablewidth{0pc}
\tablecaption{Quasar Properties}
\tabletypesize{\footnotesize}

\tablehead{
& & \multicolumn{2}{c}{Power-law fit} \\
& & \multicolumn{2}{c}{\hrulefill} \\
\colhead{Quasar} &
\colhead {Redshift} &
\colhead{$F_\lambda$(1800\,\AA)} &
\colhead{$\alpha_{\mathrm{UV}}$} &
\multicolumn{2}{c}{$\log \lambda L_\lambda$(3000\,\AA)} &
\multicolumn{2}{c}{W$_\mathrm{e}$(\ion{C}{4})} \\
& & \colhead{($10^{-15}$\,erg cm$^{-2}$\ s$^{-1}$\ \AA$^{-1}$)}
& & \multicolumn{2}{c}{(log erg s$^{-1}$)} &
\multicolumn{2}{c}{(\AA)}
}
\startdata
HE\,0143-3535 & 0.446 & $ 1.51 \pm 0.02$ & $-0.51 \pm  0.04$ & 45.477 &  0.006 & \multicolumn{2}{c}{\nodata} \\
PKS\,0252-549 & 0.539 & $ 1.36 \pm 0.02$ & $-1.45 \pm  0.04$ & 45.260 &  0.006 & 107 & 5 \\
HE\,0354-5500 & 0.267 & $ 7.49 \pm 0.05$ & $-1.40 \pm  0.02$ & 45.103 &  0.003 &  33 & 2 \\
HE\,0409-5004 & 0.817 & $ 3.53 \pm 0.03$ & $-0.60 \pm  0.02$ & 46.493 &  0.003 & \multicolumn{2}{c}{\nodata} \\
HE\,0436-2614 & 0.690 & $ 3.02 \pm 0.06$ & $-0.65 \pm  0.05$ & 46.209 &  0.009 & \multicolumn{2}{c}{\nodata} \\
HE\,0441-2826 & 0.155 & $15.36 \pm 0.04$ & $-1.13 \pm  0.01$ & 45.207 &  0.001 & 3.5 & 0.9 \\
HE\,0502-2948 & 0.552 & $ 9.86 \pm 0.05$ & $-1.53 \pm  0.01$ & 46.107 &  0.002 &  35 & 1 \\
HE\,1006-1211 & 0.693 & $ 4.11 \pm 0.03$ & $-1.04 \pm  0.02$ & 46.172 &  0.003 &  36 & 2 \\
HE\,1101-0959 & 0.186 & $ 1.02 \pm 0.02$ & $-0.33 \pm  0.07$ & 44.439 &  0.009 & 140 & 7 \\
HE\,1105-0746 & 0.340 & $ 3.38 \pm 0.03$ & $-1.35 \pm  0.03$ & 45.236 &  0.004 & \multicolumn{2}{c}{\nodata} \\
 PG\,1435-067 & 0.126 & $ 6.24 \pm 0.15$ & $-0.49 \pm  0.06$ & 44.794 &  0.010 & 168 & 8 \\
 PG\,1552+085 & 0.119 & $ 2.92 \pm 0.02$ & $ 0.11 \pm  0.02$ & 44.573 &  0.003 &  51 & 4 \\
 PG\,2233+134 & 0.325 & $ 6.14 \pm 0.04$ & $-1.33 \pm  0.02$ & 45.458 &  0.003 &   7 & 1 \\
HE\,2327-5522 & 0.494 & $ 2.72 \pm 0.04$ & $-1.14 \pm  0.03$ & 45.595 &  0.006 & 119 & 5
\enddata

\tablecomments{For the power-law fit, we adopt the convention $F_\lambda \sim \lambda^\alpha$.
The rest-frame luminosity at 3000\,\AA\ was computed assuming a $\Omega_M = 0.27$, $\Omega_\Lambda = 0.73$,
$H_\mathrm{o} = 71$\,\kms~Mpc$^{-1}$\ cosmology. The quoted {$1\sigma$} error in both flux and
luminosity reflect {\it statistical fitting uncertainties only}. The error associated with the
uncertainty in absolute flux calibration, which dominates the error in both flux and luminosity, is 4\%,
or 0.02 dex. For quasars that do not show intrinsic/associated absorption, we list the
\ion{C}{4} emission line equivalent width [W$_\mathrm{e}$(\ion{C}{4})].}

\label{tab:qsoprops}
\end{deluxetable}

%% file: tab2.tex
\begin{deluxetable}{lrrr@{\,$\pm$\,}lr@{\,$\pm$\,}l}
\tablewidth{0pc}
\tablecaption{Properties of Absorbed Quasars}

\tablehead{
\colhead{Quasar} &
\colhead{BI} &
\colhead{AI} &
\multicolumn{2}{c}{$W_\mathrm{a}$(\ion{C}{4})} &
\multicolumn{2}{c}{$v_\mathrm{max}$} \\
&
\colhead{(\kms)} &
\colhead{(\kms)} &
\multicolumn{2}{c}{(\AA)} &
\multicolumn{2}{c}{(\kms)}
}
\startdata
HE 0143-3535 &    2829 &    3701 &  36   & 4   & 13519 & 207 \\
HE 0409-5004 & \nodata & \nodata &   1.4 & 0.3 &  5417 & 165 \\
HE 0436-2614 &   11592 &   12740 & 118   & 3   & 24595 & 177 \\
HE 1105-0746 & \nodata & \nodata &   8   & 1   &  5151 & 224
\enddata
\tablecomments{For the BAL QSOs, we report balnicity and intrinsic absorption indices as defined
by \citet{weymann91} and \citet{hallai}. We also provide an
estimate of the observed \ion{C}{4} absorption equivalent width ($W_\mathrm{a}$). The quoted $1 \sigma$\ confidence
error includes both statistical and effective-continuum fitting uncertainties.}
\label{tab:absprops}
\end{deluxetable}

% All undetected in RASS .... assume <0.05 cps

% Redshifts for vmax calculation
%0.38220157033569
%0.7844612208727
%0.31717138051575
%0.556648865395

%% file: ms.bbl
\begin{thebibliography}{}

\bibitem[\protect\citeauthoryear{{Arav} et~al.}{{Arav} et~al.}{1999}]{arav99b}
{Arav}, N., {Becker}, R.~H., {Laurent-Muehleisen}, S.~A., {Gregg},
M.~D.,
  {White}, R.~L., {Brotherton}, M.~S.,  \& {de Kool}, M. 1999, \apj, 524, 566

\bibitem[\protect\citeauthoryear{{Arav}, {Li}, \& {Begelman}}{{Arav}
  et~al.}{1994}]{alb94}
{Arav}, N., {Li}, Z.-Y.,  \& {Begelman}, M.~C. 1994, \apj, 432, 62

\bibitem[\protect\citeauthoryear{{Baskin} \& {Laor}}{{Baskin} \&
  {Laor}}{2005}]{bl05}
{Baskin}, A.,  \& {Laor}, A. 2005, \mnras, 356, 1029

\bibitem[\protect\citeauthoryear{{Becker} et~al.}{{Becker}
  et~al.}{2000}]{becker00}
{Becker}, R.~H., {White}, R.~L., {Gregg}, M.~D., {Brotherton},
M.~S.,
  {Laurent-Muehleisen}, S.~A.,  \& {Arav}, N. 2000, \apj, 538, 72

\bibitem[\protect\citeauthoryear{{Becker} et~al.}{{Becker}
  et~al.}{2001}]{third}
{Becker}, R.~H., et~al. 2001, \apjs, 135, 227

\bibitem[\protect\citeauthoryear{{Boroson} \& {Green}}{{Boroson} \&
  {Green}}{1992}]{bg92}
{Boroson}, T.~A.,  \& {Green}, R.~F. 1992, \apjs, 80, 109

\bibitem[\protect\citeauthoryear{{Brandt}, {Laor}, \& {Wills}}{{Brandt}
  et~al.}{2000}]{brandt}
{Brandt}, W.~N., {Laor}, A.,  \& {Wills}, B.~J. 2000, \apj, 528, 637

\bibitem[\protect\citeauthoryear{{Brotherton} et~al.}{{Brotherton}
  et~al.}{1994}]{bro94}
{Brotherton}, M.~S., {Wills}, B.~J., {Steidel}, C.~C.,  \&
{Sargent}, W.~L.~W.
  1994, \apj, 423, 131

\bibitem[\protect\citeauthoryear{{Castor}, {Abbott}, \& {Klein}}{{Castor}
  et~al.}{1975}]{cak75}
{Castor}, J.~I., {Abbott}, D.~C.,  \& {Klein}, R.~I. 1975, \apj,
195, 157

\bibitem[\protect\citeauthoryear{{Condon} et~al.}{{Condon} et~al.}{1998}]{nvss}
{Condon}, J.~J., {Cotton}, W.~D., {Greisen}, E.~W., {Yin}, Q.~F.,
{Perley},
  R.~A., {Taylor}, G.~B.,  \& {Broderick}, J.~J. 1998, \aj, 115, 1693

\bibitem[\protect\citeauthoryear{{Crenshaw} et~al.}{{Crenshaw}
  et~al.}{1999}]{cren99}
{Crenshaw}, D.~M., {Kraemer}, S.~B., {Boggess}, A., {Maran}, S.~P.,
  {Mushotzky}, R.~F.,  \& {Wu}, C. 1999, \apj, 516, 750

\bibitem[\protect\citeauthoryear{{Everett}}{{Everett}}{2005}]{everett05}
{Everett}, J.~E. 2005, \apj, 631, 689

\bibitem[\protect\citeauthoryear{{Gallagher} et~al.}{{Gallagher}
  et~al.}{2006}]{gallsc06}
{Gallagher}, S.~C., {Brandt}, W.~N., {Chartas}, G., {Priddey}, R.,
{Garmire},
  G.~P.,  \& {Sambruna}, R.~M. 2006, \apj, 644, 709

\bibitem[\protect\citeauthoryear{{Ganguly} et~al.}{{Ganguly}
  et~al.}{2001}]{gan01a}
{Ganguly}, R., {Bond}, N.~A., {Charlton}, J.~C., {Eracleous}, M.,
{Brandt},
  W.~N.,  \& {Churchill}, C.~W. 2001, \apj, 549, 133

\bibitem[\protect\citeauthoryear{{Ganguly} et~al.}{{Ganguly}
  et~al.}{2006}]{gan06a}
{Ganguly}, R., {Sembach}, K.~R., {Tripp}, T.~M., {Savage}, B.~D.,
\& {Wakker},
  B.~P. 2006, \apj, 645, 868

\bibitem[\protect\citeauthoryear{{Gaskell} et~al.}{{Gaskell}
 et~al.}{2004}]{gaskell04}
{Gaskell}, C.~M., {Goosmann}, R.~W., {Antonucci}, R.~R.~J., \& {Whysong},
D.~H. 2004, \apj, 616, 147

\bibitem[\protect\citeauthoryear{{Gaskell} \& {Benker}}{{Gaskell} \&
{Benker}}{2006}]{gaskell06}
{Gaskell}, C.~M., \& {Benker}, A.~J. 2006, \apj, submitted

\bibitem[\protect\citeauthoryear{{Hall} et~al.}{{Hall} et~al.}{2002}]{hallai}
{Hall}, P.~B., et~al. 2002, \apjs, 141, 267

\bibitem[\protect\citeauthoryear{{Hall} et~al.}{{Hall} et~al.}{2004}]{hall04}
{Hall}, P.~B., et~al. 2004, \aj, 127, 3146

\bibitem[\protect\citeauthoryear{{Hamann}}{{Hamann}}{1998}]{ham98}
{Hamann}, F. 1998, \apj, 500, 798

\bibitem[\protect\citeauthoryear{{Hamann} et~al.}{{Hamann}
  et~al.}{2001}]{ham01}
{Hamann}, F.~W., {Barlow}, T.~A., {Chaffee}, F.~C., {Foltz}, C.~B.,
\&
  {Weymann}, R.~J. 2001, \apj, 550, 142

\bibitem[\protect\citeauthoryear{{Hewett} \& {Foltz}}{{Hewett} \&
  {Foltz}}{2003}]{hf03}
{Hewett}, P.~C.,  \& {Foltz}, C.~B. 2003, \aj, 125, 1784

\bibitem[\protect\citeauthoryear{{Hewett}, {Foltz}, \& {Chaffee}}{{Hewett}
  et~al.}{1995}]{lbqs6}
{Hewett}, P.~C., {Foltz}, C.~B.,  \& {Chaffee}, F.~H. 1995, \aj,
109, 1498

\bibitem[\protect\citeauthoryear{{Kellermann} et~al.}{{Kellermann}
  et~al.}{1994}]{kel94}
{Kellermann}, K.~I., {Sramek}, R.~A., {Schmidt}, M., {Green}, R.~F.,
\&
  {Shaffer}, D.~B. 1994, \aj, 108, 1163

\bibitem[\protect\citeauthoryear{{Kim-Quijano} et~al.}{{Kim-Quijano}
  et~al.}{2003}]{stis}
{Kim-Quijano}, J., et~al. 2003, {STIS Intrument Handbook, Version
7.0}
  (Baltimore: STScI)

\bibitem[\protect\citeauthoryear{{Kimble} et~al.}{{Kimble}
  et~al.}{1998}]{kimble98}
{Kimble}, R.~A., et~al. 1998, in Proc. SPIE Vol. 3356, p. 188-202,
Space
  Telescopes and Instruments V, Pierre Y. Bely; James B. Breckinridge; Eds.,
  188

\bibitem[\protect\citeauthoryear{{Kriss}}{{Kriss}}{2002}]{kriss02}
{Kriss}, G.~A. 2002, in ASP Conf. Ser. 255: Mass Outflow in Active
Galactic
  Nuclei: New Perspectives, ed. D.~M. {Crenshaw}, S.~B. {Kraemer}, \& I.~M.
  {George} (San Francisco: ASP), 69

\bibitem[\protect\citeauthoryear{{Krolik} \& {Kriss}}{{Krolik} \&
  {Kriss}}{2001}]{kk01}
{Krolik}, J.~H.,  \& {Kriss}, G.~A. 2001, \apj, 561, 684

\bibitem[\protect\citeauthoryear{{Laor} \& {Brandt}}{{Laor} \&
  {Brandt}}{2002}]{lb02}
{Laor}, A.,  \& {Brandt}, W.~N. 2002, \apj, 569, 641

\bibitem[\protect\citeauthoryear{{Lynds}}{{Lynds}}{1967}]{lynds67}
{Lynds}, C.~R. 1967, \apj, 147, 396

\bibitem[\protect\citeauthoryear{{McDowell} et~al.}{{McDowell}
  et~al.}{1995}]{mcdowell95}
{McDowell}, J.~C., {Canizares}, C., {Elvis}, M., {Lawrence}, A.,
{Markoff}, S.,
  {Mathur}, S.,  \& {Wilkes}, B.~J. 1995, \apj, 450, 585

\bibitem[\protect\citeauthoryear{{Murray} et~al.}{{Murray}
  et~al.}{1995}]{mur95}
{Murray}, N., {Chiang}, J., {Grossman}, S.~A.,  \& {Voit}, G.~M.
1995, \apj,
  451, 498

\bibitem[\protect\citeauthoryear{{Neugebauer} et~al.}{{Neugebauer}
  et~al.}{1987}]{neu87}
{Neugebauer}, G., {Green}, R.~F., {Matthews}, K., {Schmidt}, M.,
{Soifer},
  B.~T.,  \& {Bennett}, J. 1987, \apjs, 63, 615

\bibitem[\protect\citeauthoryear{{Press} et~al.}{{Press}
  et~al.}{1992}]{nrpress}
{Press}, W.~H., {Teukolsky}, S.~A., {Vetterling}, W.~T.,  \&
{Flannery}, B.~P.
  1992, {Numerical recipes in C. The art of scientific computing} (Cambridge:
  University Press, 1992, 2nd ed.)

\bibitem[\protect\citeauthoryear{{Proga} \& {Kallman}}{{Proga} \&
  {Kallman}}{2004}]{pk04}
{Proga}, D.,  \& {Kallman}, T.~R. 2004, \apj, 616, 688

\bibitem[\protect\citeauthoryear{{Proga}, {Stone}, \& {Drew}}{{Proga}
  et~al.}{1998}]{psd98}
{Proga}, D., {Stone}, J.~M.,  \& {Drew}, J.~E. 1998, \mnras, 295,
595

\bibitem[\protect\citeauthoryear{{Richards}}{{Richards}}{2001}]{rich01b}
{Richards}, G.~T. 2001, \apjs, 133, 53

\bibitem[\protect\citeauthoryear{{Richards} et~al.}{{Richards}
  et~al.}{2002}]{rich02b}
{Richards}, G.~T., {Vanden Berk}, D.~E., {Reichard}, T.~A., {Hall},
P.~B.,
  {Schneider}, D.~P., {SubbaRao}, M., {Thakar}, A.~R.,  \& {York}, D.~G. 2002,
  \aj, 124, 1

\bibitem[\protect\citeauthoryear{{Schmidt} \& {Green}}{{Schmidt} \&
  {Green}}{1983}]{sg83}
{Schmidt}, M.,  \& {Green}, R.~F. 1983, \apj, 269, 352

\bibitem[\protect\citeauthoryear{{Scott} et~al.}{{Scott}
  et~al.}{2004}]{scott04b}
{Scott}, J.~E., {Kriss}, G.~A., {Brotherton}, M., {Green}, R.~F.,
{Hutchings},
  J., {Shull}, J.~M.,  \& {Zheng}, W. 2004, \apj, 615, 135

\bibitem[\protect\citeauthoryear{{Shang} et~al.}{{Shang}
  et~al.}{2005}]{shang05}
{Shang}, Z., et~al. 2005, \apj, 619, 41

\bibitem[\protect\citeauthoryear{{Sulentic} et~al.}{{Sulentic}
  et~al.}{2006}]{sulentic06}
{Sulentic}, J.~W., {Dultzin-Hacyan}, D., {Marziani}, P., {Bongardo},
C.,
  {Braito}, V., {Calvani}, M.,  \& {Zamanov}, R. 2006, Revista Mexicana de
  Astronomia y Astrofisica, 42, 23

\bibitem[\protect\citeauthoryear{{Telfer} et~al.}{{Telfer}
  et~al.}{2002}]{telfer02}
{Telfer}, R.~C., {Zheng}, W., {Kriss}, G.~A.,  \& {Davidsen}, A.~F.
2002, \apj,
  565, 773

\bibitem[\protect\citeauthoryear{{Turnshek} et~al.}{{Turnshek}
  et~al.}{1988}]{turn88}
{Turnshek}, D.~A., {Grillmair}, C.~J., {Foltz}, C.~B.,  \&
{Weymann}, R.~J.
  1988, \apj, 325, 651

\bibitem[\protect\citeauthoryear{{Turnshek} et~al.}{{Turnshek}
  et~al.}{1997}]{turn97}
{Turnshek}, D.~A., {Monier}, E.~M., {Sirola}, C.~J.,  \& {Espey},
B.~R. 1997,
  \apj, 476, 40

\bibitem[\protect\citeauthoryear{{Vestergaard}}{{Vestergaard}}{2003}]{ves03}
{Vestergaard}, M. 2003, \apj, 599, 116

\bibitem[\protect\citeauthoryear{{Vestergaard} \& {Wilkes}}{{Vestergaard} \&
  {Wilkes}}{2001}]{vw01}
{Vestergaard}, M.,  \& {Wilkes}, B.~J. 2001, \apjs, 134, 1

\bibitem[\protect\citeauthoryear{{Weymann} et~al.}{{Weymann}
  et~al.}{1991}]{weymann91}
{Weymann}, R.~J., {Morris}, S.~L., {Foltz}, C.~B.,  \& {Hewett},
P.~C. 1991,
  \apj, 373, 23

\bibitem[\protect\citeauthoryear{{Weymann}, {Turnshek}, \&
  {Christiansen}}{{Weymann} et~al.}{1985}]{weymann85}
{Weymann}, R.~J., {Turnshek}, D.~A.,  \& {Christiansen}, W.~A. 1985,
in
  Astrophysics of Active Galaxies and Quasi-Stellar Objects, ed. J.~S.
  {Miller}, 333

\bibitem[\protect\citeauthoryear{{White} et~al.}{{White} et~al.}{2000}]{second}
{White}, R.~L., et~al. 2000, \apjs, 126, 133

\bibitem[\protect\citeauthoryear{{Wills}, {Brandt}, \& {Laor}}{{Wills}
  et~al.}{1999}]{wbl99}
{Wills}, B.~J., {Brandt}, W.~N.,  \& {Laor}, A. 1999, \apjl, 520,
L91

\bibitem[\protect\citeauthoryear{{Wills} et~al.}{{Wills}
  et~al.}{1993}]{wills93}
{Wills}, B.~J., {Brotherton}, M.~S., {Fang}, D., {Steidel}, C.~C.,
\&
  {Sargent}, W.~L.~W. 1993, \apj, 415, 563

\bibitem[\protect\citeauthoryear{{Wise} et~al.}{{Wise} et~al.}{2004}]{wise04}
{Wise}, J.~H., {Eracleous}, M., {Charlton}, J.~C.,  \& {Ganguly}, R.
2004,
  \apj, 613, 129

\bibitem[\protect\citeauthoryear{{Woodgate} et~al.}{{Woodgate}
  et~al.}{1998}]{stis98}
{Woodgate}, B.~E., et~al. 1998, \pasp, 110, 1183

\end{thebibliography}
